\documentclass[twocolumn,superscriptaddress]{revtex4}

\usepackage{natbib}
\usepackage{graphicx}

\begin{document}

\title{Hubbard exciton revealed by time-domain optical spectroscopy}

\author{Fabio Novelli}
\affiliation{Department of Physics, Universit\`a degli Studi di Trieste, 34127 Trieste, Italy}
\author{Daniele Fausti}
\email{daniele.fausti@elettra.trieste.it}
\affiliation{Department of Physics, Universit\`a degli Studi di Trieste, 34127 Trieste, Italy}
\affiliation{Sincrotrone Trieste S.C.p.A., 34127 Basovizza, Italy}
\author{Julia Reul}
\affiliation{Department of Physics, University of Cologne, 50923 K\"oln, Germany}
\author{Federico Cilento}
\affiliation{Sincrotrone Trieste S.C.p.A., 34127 Basovizza, Italy}
\author{Paul H. M. van Loosdrecht}
\affiliation{Zernike Institute for Advanced Materials, University of Groningen, 9747 AG Groningen, The Netherlands}
\author{Agung A. Nugroho}
\affiliation{Faculty of Mathematics and Natural Sciences, Jl. Ganesa 10 Bandung, 40132, Indonesia}
\author{Thomas T. M. Palstra}
\affiliation{Zernike Institute for Advanced Materials, University of Groningen, 9747 AG Groningen, The Netherlands}
\author{Markus Gr\"uninger}
\affiliation{Department of Physics, University of Cologne, 50923 K\"oln, Germany}
\author{Fulvio Parmigiani}
\affiliation{Department of Physics, Universit\`a degli Studi di Trieste, 34127 Trieste, Italy}
\affiliation{Sincrotrone Trieste S.C.p.A., 34127 Basovizza, Italy}
\date{August 28, 2012}

\begin{abstract}
We use broadband ultra-fast pump-probe spectroscopy in the visible range to study the lowest excitations across the Mott-Hubbard gap in the orbitally ordered insulator YVO$_{3}$. Separating thermal and non-thermal contributions to the optical transients, we show that the total spectral weight of the two lowest peaks is conserved, demonstrating that both excitations correspond to the same multiplet. The pump-induced transfer of spectral weight between the two peaks reveals that the low-energy one is a Hubbard exciton, i.e. a resonance or bound state between a doublon and a holon. Finally, we speculate that the pump-driven spin-disorder can be used to quantify the kinetic energy gain of the excitons in the ferromagnetic phase.
\end{abstract}

\maketitle

The physical properties of materials characterized by strong electron-electron interactions are determined by the competitive minimization of the potential energy and kinetic energy of the electrons. While the potential energy is lowest for localized electrons, the tendency to a metallic state increases it at the expense of kinetic energy. The fine tuning of the system's parameter results in the exotic ordering phenomena characterizing transition metal oxides' (TMOs) phase diagrams\cite{yvo:ima98}. In a Mott-Hubbard insulator the lowest electronic excitation across the gap creates, in the most simple case, an empty site (holon in the lower Hubbard band) and a doubly occupied site (doublon in the upper Hubbard band)\cite{yvo:fuj94}. In the single-band Hubbard model, the energy of this transition is solely determined by the on-site Coulomb repulsion between electrons leading to an effective energy cost $U$. Typically, holon and doublon are not bound to each other, but more composite excited states have been predicted by the extended Hubbard model including non-local interactions\cite{yvo:bri95,yvo:neu98,yvo:hub03,yvo:wal11,yvo:lei09}. In particular a new kind of
bound state between a holon and a doublon was recently introduced and named Hubbard exciton (HE)\cite{yvo:ess01,yvo:mat05,yvo:gos08,yvo:mat09}. While the formation of HEs can be driven by a drop of Coulomb energy\cite{yvo:gal97,yvo:jec03}, as in simple semiconductors, a kinetic energy loss could further stabilize the excitonic state in magnetic environments\cite{yvo:zha98,yvo:cla93,yvo:wro02}. Non-localized HEs have been extensively studied in the framework of high-temperature super-conductivity\cite{yvo:wan96,yvo:col06} in relation with the proposed kinetic energy driven formation of the condensate\cite{yvo:hir02,yvo:mol02,yvo:phi10}.

In this letter we show that an extended HE picture rationalize the optical properties of YVO$_3$, a case-study for Mott-insulators TMOs. In this scenario, the optical transitions (see FIG. 1) observed at 2.4\,eV reflects the ``single-particle'' band (SP) whereas the one at 1.8\,eV is attributed to a kinetic energy related Hubbard exciton\cite{yvo:reu12}. Our pump-probe spectroscopic measurements in the 1.65-2.75\,eV range revealed that the spectral weight ($SW$) is directly transferred between the two peaks, confirming the excitonic nature of the low-energy feature. Together with this, we measured both thermal and non-thermal effects and quantified the kinetic energy contribution to the formation of the HE. Our study provides a new methodology, based on both static and time-domain spectroscopy, that can be used to unravel the complex nature of high-energy excitations in insulating TMOs and, more broadly, to study the kinetic energy-based mechanisms in strongly-correlated materials.

\begin{figure}[t!] \centering
\includegraphics[width=8.4cm]{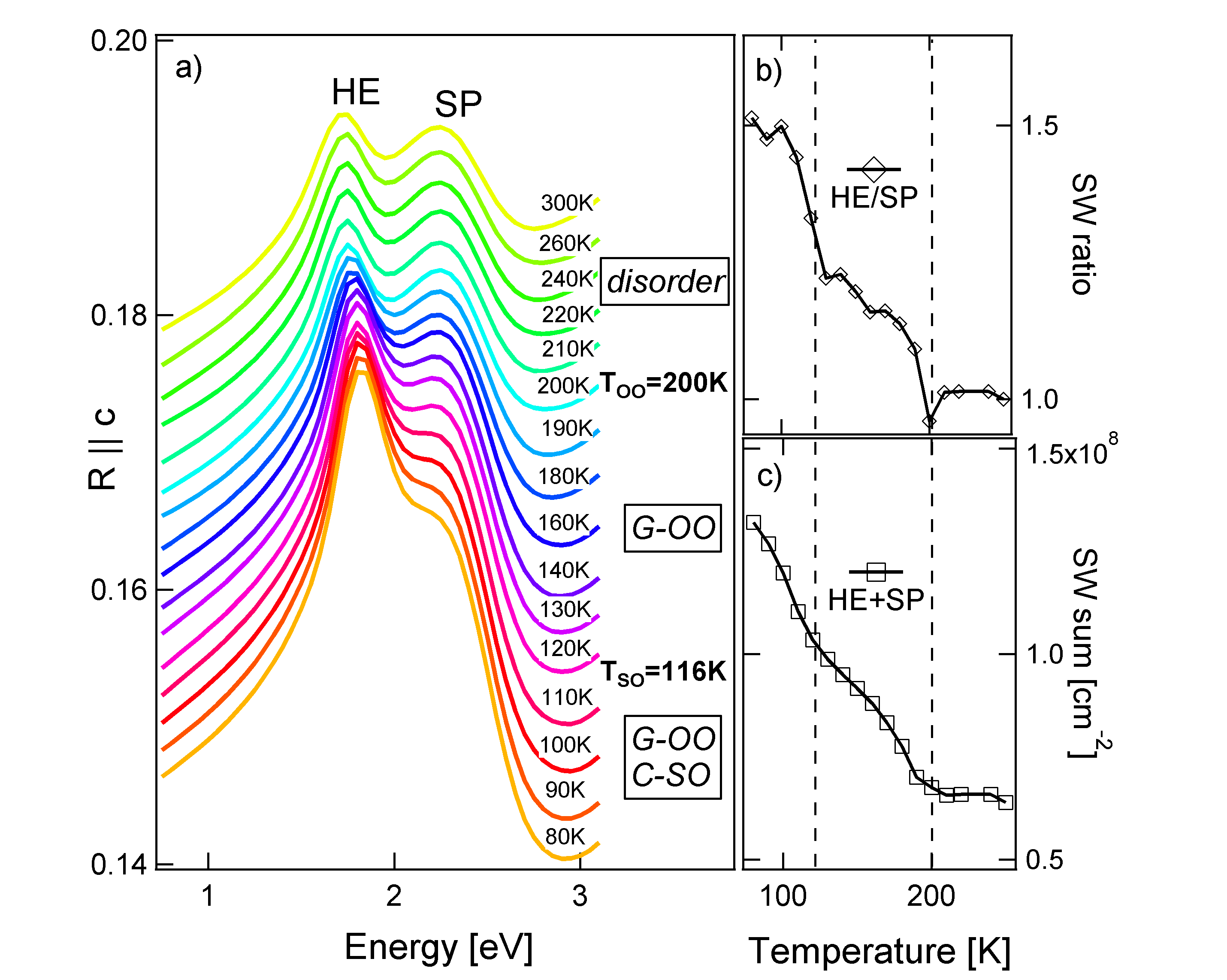}
\caption{\label{FIG1} \textbf{Reflectivity as a function of temperature of YVO$_3$ obtained by ellipsometry data (a).} The ratio  between the spectral weight of the excitonic and single particle band (b) increases rapidly while entering the G-OO phase and further increases at the spin ordering temperature. Both the SP and the HE spectral weights raise upon cooling (c). In (b) and (c) the dashed lines represent the transitions temperatures towards the orbital ordering (T$_{OO}$=200\,K) and the additional spin ordering (T$_{SO}$=116\,K). In (b) the ratio is normalized to the value at 300\,K. Note that the reflectivity measurements in (a) are displaced for clarity from the measurement at 80\,K.}
\end{figure}

The observation of multiple temperature-induced magnetization reversals\cite{yvo:ren98} accompanied by a series of structural\cite{yvo:bla02}, magnetic\cite{yvo:ulr03}, and orbital transitions\cite{yvo:nog00} turned the Mott insulator YVO$_3$ into a case-study among TMOs. At room temperature YVO$_3$ has a $Pbnm$ orthorhombic unit cell, where each VO$_6$ octahedron is tilted and distorted. The V$^{3+}$ ions have a 3d$_2$ electronic configuration\cite{yvo:pen99} so that two electrons occupy the $t_{2g}$ orbitals. At 200\,K a phase transition to a monoclinic phase ($Pb11$\cite{yvo:tsv04} or $P2_1/c$\cite{yvo:bla02,yvo:ree06,yvo:bea10}) occurs with G-type orbital order (G-OO). In this state, the $d_{yz}$ or $d_{zx}$ orbitals are alternately occupied both in the $ab$ plane and along the $c$ axis (FIG. 2 (e)). While still monoclinic and G-OO, at T$_{SO}$=116\,K a C-type spin order emerges, characterized by an antiferromagentic (AFM) spin configuration in the $ab$ plane and ferromagentic (FM) along the $c$ axis (FIG. 2 (f)). For further cooling below T$_{N\acute{e}el,2}$=77\,K a structural, orbital, and magnetic transition occurs: the system recovers the
$Pbnm$ crystalline structure while the orbital ordering switches to C-type (C-OO) and the spin order to G-type (G-SO)\cite{yvo:saw96,yvo:zho03,yvo:zho04,yvo:der07}. The low-temperature phase will not be further discussed in the following (see additional materials).

\section{\label{results} Results}

All phase transitions can be identified by monitoring the optical properties of YVO$_3$. The region of the optical spectra that is
particularly sensitive to the orbital physics is the visible range.
The two peaks characterizing the optical properties with energies of
1.8\,eV (HE) and 2.4\,eV (SP) are largely debated\cite{yvo:mos07,yvo:mos09,yvo:ots06,yvo:zho03,yvo:der07} and commonly assigned to
d$_i^2$d$_j^2$-d$_i^1$d$_j^3$ transitions between two different
V$_i$ and V$_j$ sites\cite{yvo:miy02,yvo:tsv04}.
Multiplet calculations\cite{yvo:tsv04} account for the presence of the different peaks but fail in reproducing their temperature
dependence: both the 1.8\,eV and 2.4\,eV excitations gain spectral weight with decreasing temperature and approaching the spin ordering transition at T$_{SO}$=116\,K (FIG. 1 (c)),
indicating that they correspond to the same high-spin state\cite{yvo:reu12}. 
The SW gain with decreasing temperature is much stronger for the lower peak, which tentatively has been attributed to excitonic behavior. In fact from Hund's rules we expect the kinetic energy contribution to the formation of the Hubbard exciton to be more relevant in a FM-ordered phase\cite{yvo:gos08}.

In order to distinguishing the effects of temperature and orbital
disorder we performed pump-probe reflectivity measurements.

The complex dielectric function $\varepsilon(\omega)$ was measured by ellipsometry, for details see Ref.\cite{yvo:reu12}. The static normal-incidence reflectivity
$R(\omega)$ reported in FIG. 1 was calculated from $\varepsilon(\omega)$. Broadband super-continuum probe experiments combined with
an ultrafast optical pump at 1.55\,eV were performed on freshly polished $ac$ oriented YVO$_3$ samples mounted on the cold finger
of a helium-flow cryostat. The reflectivity changes as a function
of pump-probe delay $\frac{\Delta{R}}{R}(\omega,t)=\frac{R(\omega,t)-R(\omega)}{R(\omega)}$ induced by 80\,fs pump pulses ($E_{pump}$=1.55\,eV,  fluence$<$4\,mJ/cm$^2$, at 40\,KHz
repetition rate and with polarization parallel to the $a$ axis) were measured as a function of energy (for
1.65\,eV\,$< E_{probe} < $ 2.75\,eV) and temperature. The linearity of the response was checked in all
phases up to 8\,mJ/cm$^2$. 

The three phases are characterized by the different responses summarized in FIG. 2 for probe
polarization parallel to the $c$ axis (see online additional material for the intermediate temperatures).

\begin{figure}[b!] \centering
\includegraphics[width=8.4cm]{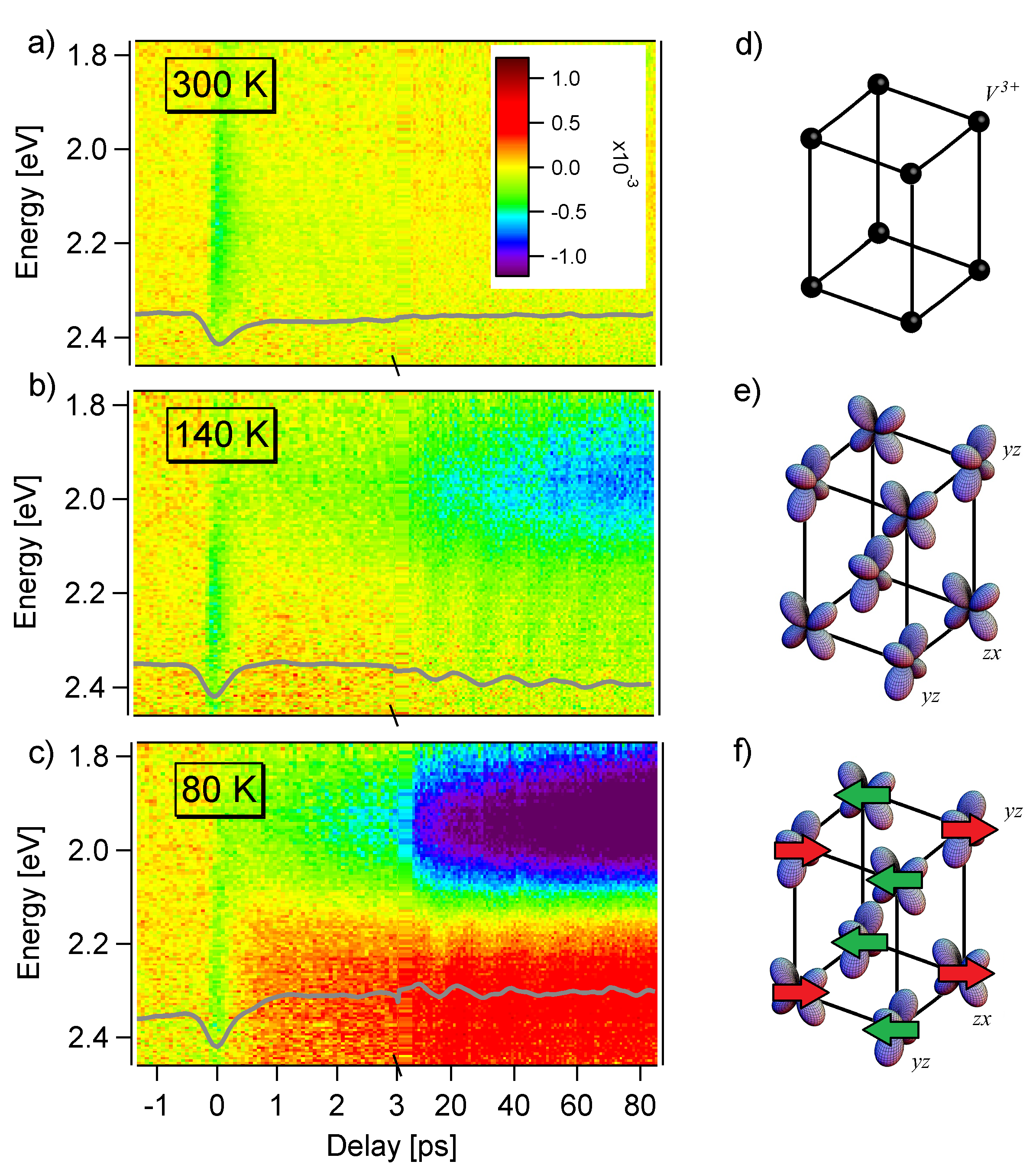}
\caption{\label{FIG2} \textbf{Broadband transient reflectivity spectra.}
Transient reflectivity as a function of wavelength (energy) and
pump-probe delay for the three different phases: (a) disordered, T=300\,K; (b) G-type orbital order, T=140\,K; (c) C-type spin order, T=80\,K.
Gray lines report the transient reflectivity at the fixed energy of 2.33\,eV as retrieved in standard
single-color pump-probe measurements.
The oscillatory trend has been assigned to acoustic
vibrations\cite{yvo:maz08} and will be ignored. For each panel, the respective ordering patterns are sketched on the right hand side (d, e, f).}
\end{figure}

At room temperature, the pump-probe measurements are solely characterized by a very fast negative variation
of the reflectivity (decay time $\tau$ $\approx$ 0.5\,ps) extending over the investigated spectral region (FIG. 2 (a)). On the contrary,
the low-temperature phases are characterized by a more composite optical response with fast and slow components, confirming previously reported
single-color measurements\cite{yvo:maz08,yvo:yus10}.
The time-domain reflectivity measurements in the G-OO phase (116\,K $<$ T $<$ 200\,K) are characterized by a slow negative response
centered at 1.94\,eV (FIG. 2 (b)), which gets more pronounced as the temperature is lowered (FIG. 2(c)).
Finally, entering the S-OO phase (80\,K $<$ T $<$ 116\,K) a positive variation of the reflectivity appears at energies higher than 2.1\,eV.

\begin{figure}[t!] \centering
\includegraphics[width=8.5cm]{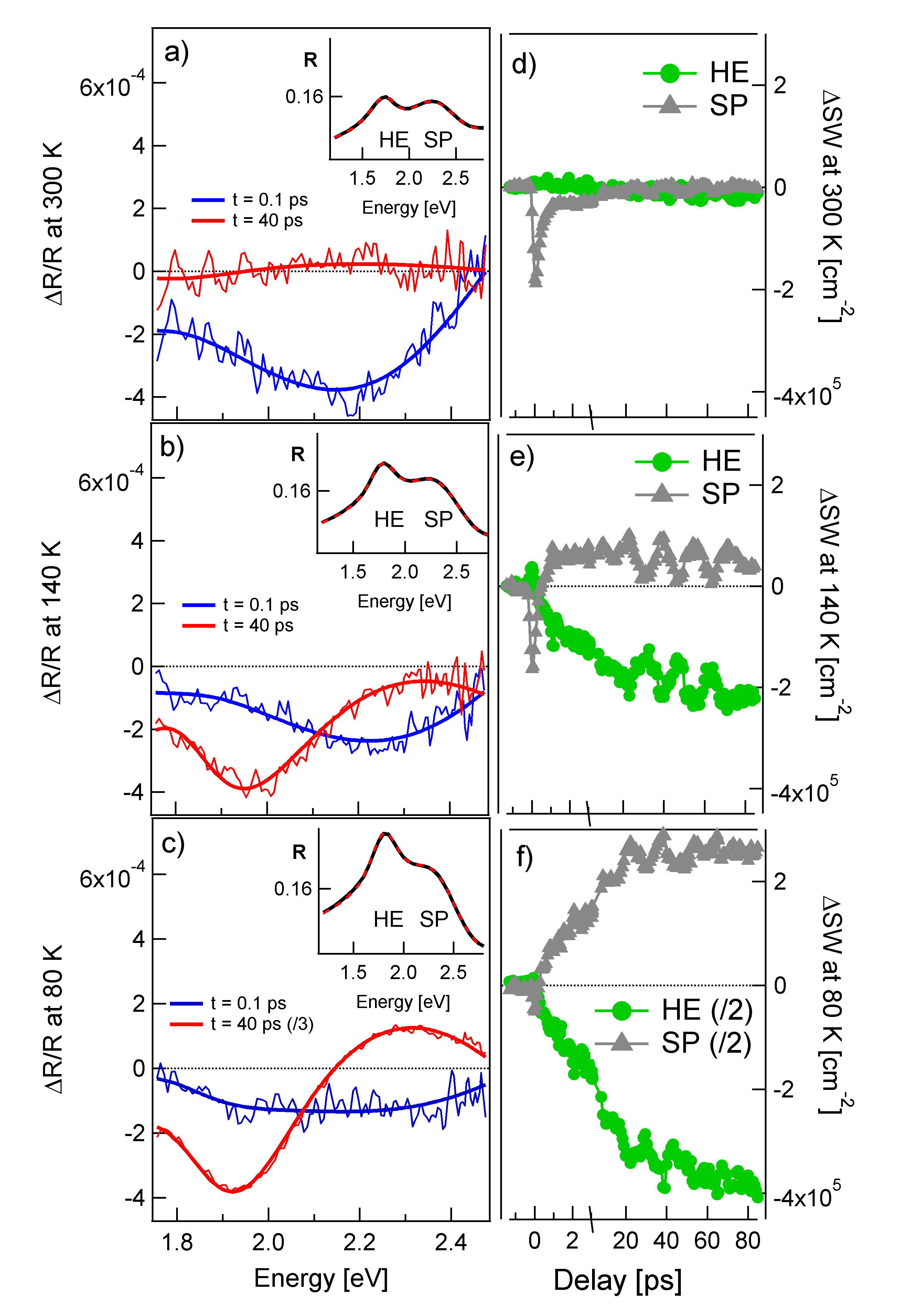}
\caption{\label{FIG3} \textbf{From transient reflectivity to spectral
weight.} The transient reflectance is fitted using a variational
approach on the static fit. (a), (b) and (c) depict the transient
reflectance for
two characteristic times, typical of the ``fast'' and ``slow'' dynamics, at 300\,K, 140\,K, and 80\,K, respectively,
and the relative variational fit. In the inserts the
static reflectance for the different phases
are reported\cite{yvo:reu12}. (d), (e) and (f) show the time evolution of the $SW$ variations
in the different phases (see text).}
\end{figure}

\section{\label{methods} Methods}

Our time-domain spectroscopic technique offers a direct view on the pump-induced changes of the reflectivity over a broad
frequency range. This has significant advantages compared to a single-color pump-probe experiment.
It allows, for instance, to determine the time-dependent spectral weight of the different features characterizing the optical response. 
The observation of a direct transfer of $SW$ between the HE and SP band in YVO$_3$ is the central experimental result of this study.

In order to calculate the pump-induced evolution of the $SW$ of the two bands
from the transient reflectivity data, we proceed as follow. 
We consider the model used to fit the static ellipsometry, we
calculate the equilibrium reflectivity ($R_0$) and we fit the
measured transient reflectance ($\Delta{R}_{exp}(t)/R_{exp}$) with a
differential model $(R(t)-R_0)/R_0$, where $R(t)$ is a model for
the perturbed reflectivity obtained by variation of the parameters
used to fit the equilibrium data.

Among the 21 parameters used to fit the features in the static $\varepsilon(\omega)$ (six Gaussian and one Tauc-Lorentz oscillators, see additional materials),
the minimal set of parameters needed to obtain good differential fits at all temperatures and times contains only the following four: the amplitude,
central frequency, and width of the oscillator describing the HE and the amplitude of the SP one. FIG. 3 (a), (b), and (c) show typical 
fits obtained for the fast (blue) and slow response (red) in the different phases. 
The obtained values for the temporal evolution of the oscillator parameters are used to calculate the time-domain evolution of the spectral weight.
The variation of the spectral weight for the two bands in time is calculated independently as the difference between the $SW$ calculated for the model at a specific time
($SW_{HE,SP}^{pumped}(t)$) minus the static spectral weight ($SW_{HE,SP}^{static}$) (each spectral weight is calculated by numerical integration of the
optical conductivity describing the band).
The time dependence of $\Delta$$SW(t)=SW^{pumped}(t)-SW^{static}$ of HE and SP
as a function of the pump-probe delay ``t'' is reported in FIG. 3 (d), (e), and (f) for three significative temperatures. 
The fast response (t $<$ 3\,ps) in all phases is entirely described by a variation of the SP peak while the HE seems to be unaffected by photo-excitation in the first few picoseconds. 
This evidence can be rationalized as ground state depletion, confirming that only the higher energy optical transition is of single-particle origin.
On the other hand the slow response (t $>$ 10\,ps), revealed in the orbitally ordered phases, is related to spectral weight changes of both HE and SP bands.

\section{\label{discussion} Discussion}

The time-domain response cannot be accounted for by photo-induced heating. The
spectral weight of both oscillators increases upon cooling (FIG. 1) and it is therefore expected that a transient laser-induced heating would result in
a decrease of the spectral weight of both SP and HE. However, the time-domain measurements (FIG. 2) reveal that only the $SW$ of the HE peak decreases,
while the SP band shows the opposite behavior, which rules out a simple heating effect.
The non-thermal $SW$ gain of the high energy oscillator (FIG. 4) lasts up to $\approx$ 400\,ps and only at longer times the measurements indicate a $SW$ loss for both oscillators. A comparison between green and black curves in FIG. 4b reveals that 1\,ns after photoexcitation the degrees of freedom have not reached the thermal equilibrium.

\begin{figure}[t!] \centering
\includegraphics[width=8.4cm]{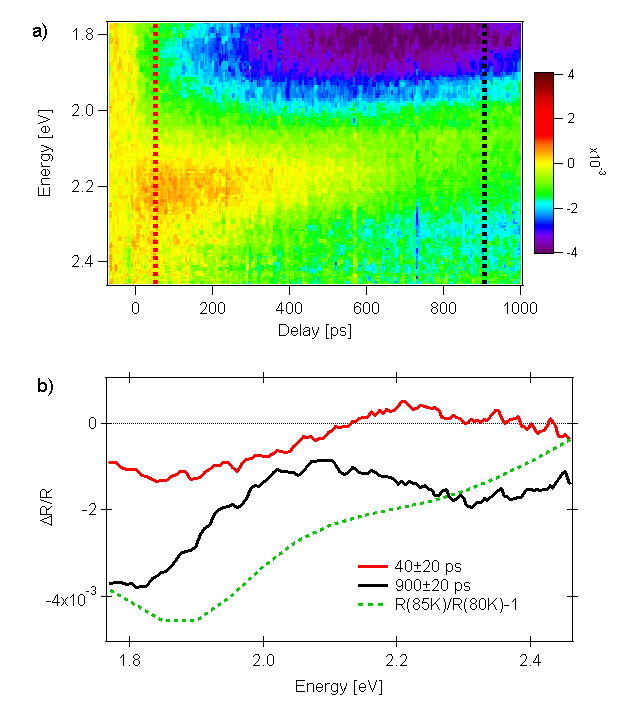}
\caption{\label{FIG4} \textbf{Long timescale pump-probe measure at 80\,K}. a) Time-domain reflectivity data at 80\,K. Note that the fast response visible in FIG. 2 is absent beacuse of the coarse temporal step used. b) Transient reflectance at t = 40\,ps (red curve) and 900\,ps (black curve). The expected thermal response $(R^{static}(85\,K)-R^{static}(80\,K))/R^{static}(80\,K)$ is shown for comparison (green curve).}
\end{figure}

Following those considerations we analyze the transient spectral weight measured at t $>$ 40\,picoseconds (well beyond the electronic relaxation) as resulting from a thermal contribution and a non-thermal one. 
The thermal contribution to the $SW$ variation of the HE and SP peaks ($SW_{HE}^{static}(T+\Delta{T})$ and $SW_{SP}^{static}(T+\Delta{T})$) can be calculated by interpolation of the temperature behaviour of the static measurements at  $T+\Delta{T}$, where $\Delta{T}$ is the photo-induced heating obtained from thermodynamic considerations (details in additional material).
The $SW$ variations of non-thermal origin can therefore be calculated subtracting the thermal contribution to the experimental values: 

\small$\Delta{SW}_{HE}^{non-thermal} = SW_{HE}^{pumped}(50\,ps) - SW_{HE}^{static}(T+\Delta{T})$

$\Delta{SW}_{SP}^{non-thermal} = SW_{SP}^{pumped}(50\,ps) - SW_{SP}^{static}(T+\Delta{T})$\normalsize, 
for HE and SP respectively, where $SW_{i}^{pumped}(50\,ps)$ ($i=HE,SP$) are the means of the measured photo-excited spectral weights at t=50$\pm$30\,ps after the pump arrival.

The non-thermal components of the pump-driven spectral weight variations at different equilibrium temperatures are reported in FIG. 5. It is evident that the non-thermal contributions consist of a direct exchange of spectral weight between the HE and SP: this result prove that a the two lowest lying optical excitations belong to the same multiplet and that the 1.8\,eV transition is of excitonic nature\cite{yvo:reu12}. Moreover, the photo-induced increase of $SW$ transfer from the HE to the SP in the C-SO phase (FIG. 5 for T $<$ 120\,K) highlighs the kinetic energy contribution to the formation of the HE.

\begin{figure}[b] \centering
\includegraphics[width=8.4cm]{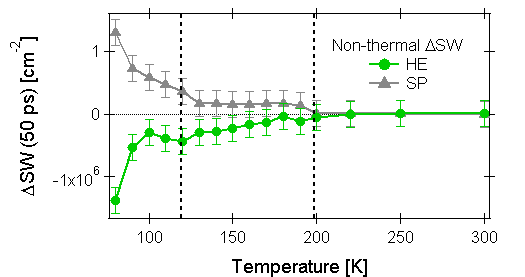}
\caption{\label{FIG5} \textbf{Non-thermal spectral weight changes of the HE and SP peaks}. The excitonic nature of the low-energy transition is revealed by the direct $SW$ exchange between the two excitations (see text). The error bars are estimated from the fitting procedure.}
\end{figure}

We argue that the spectral weight loss of the HE in favor of the SP is driven by pump-induced spin-disorder (FIG. 6). Two adjacent and c-oriented orbital-chains of YVO$_3$ in the G-OO/C-SO phase are sketched in the first row of FIG. 6. 
The spin order is FM along the $c$ axis ($J_c<0$) while it is anti-ferromagnetic in the orthogonal direction ($J_a>0$).
The photo-exctitation ($a$ axis polarized) transfers charges between chains, leaving one excess electron on a site and one vacancy on the nearest as depicted in FIG. 6 (t=0). The magnetic coupling between an excited and a non-excited site along $c$ changes as a consequence of such a charge redistribution ($J_c^*$). This dramatic perturbation of the spin coupling can be grasped by the following considerations. The Goodenough-Kanamori rules\cite{yvo:wei97}, stating that the super-exchange coupling $J$ between half-filled and empty orbitals is negative, is consistent with the FM order observed along the $c$ axis. At simplest, the pump driven changes in the orbital occupation leads to half filling, inducing a switch from FM to AFM exchange coupling. The electronic subsystem relaxes within a few picoseconds (t$_1$) while the spins remain locally perturbed (tilted blue arrows in FIG. 6 for t=t$_1$). We argue that, being the orbital angular momentum completely quenched, the coupling mechanism allowing the magnetic subsystem to transfer energy to other degrees of freedom is very weak. Therefore, the local perturbation of the spins diffuses (t$_2$) leading to spin disorder. This partially suppress the kinetic energy gain of the HE and lead, subsequently, to a non-thermal spectral weight transfer to the SP peak. A fully thermalized state is achieved only on much longer timescales ($>$ 1\,ns) through spin-lattice coupling\cite{yvo:yus10}.

\begin{figure}[h!] \centering
\includegraphics[width=7.4cm]{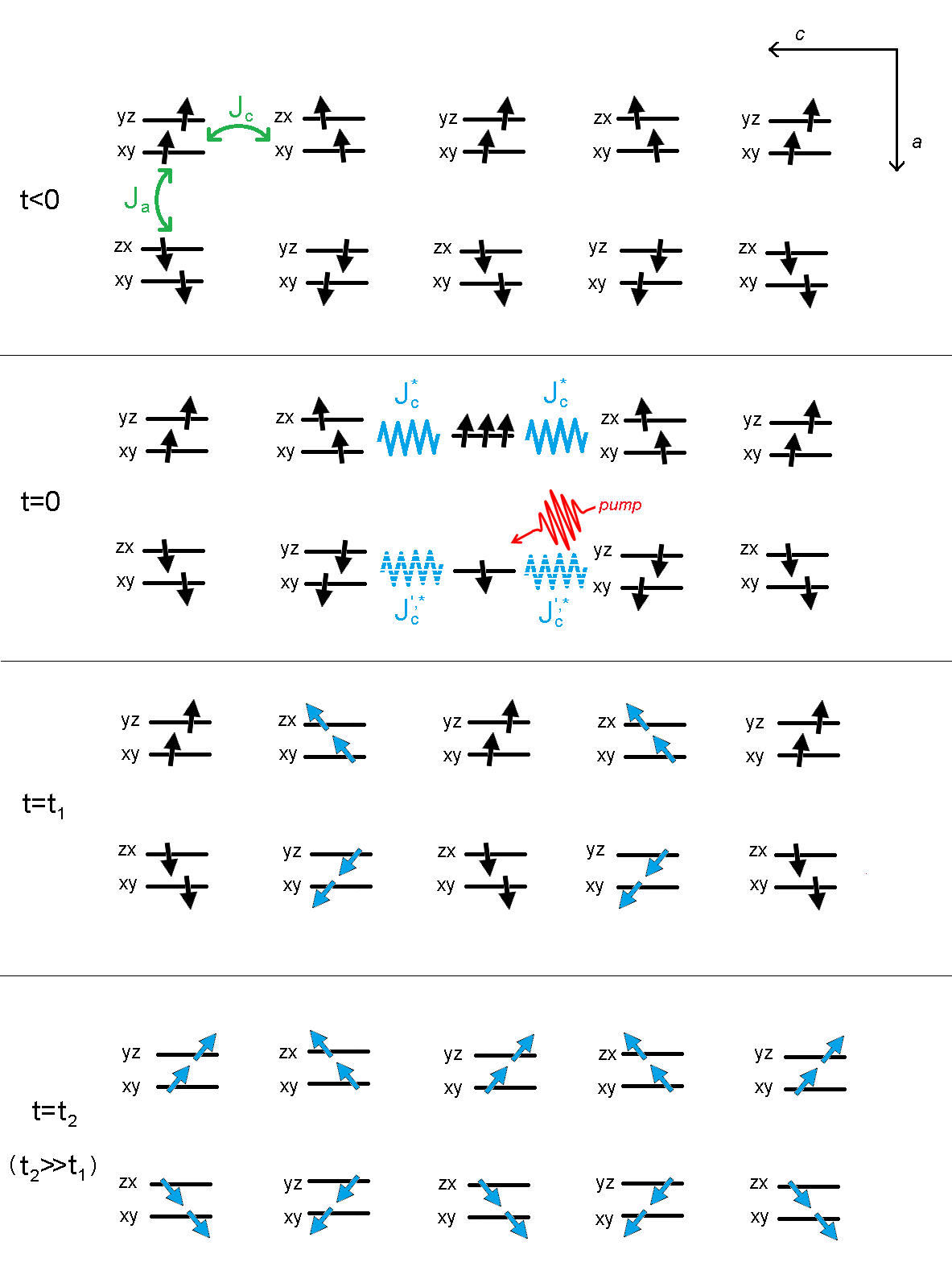}
\caption{\label{FIG6} \textbf{Non-thermal spin disorder.} YVO$_3$ is in the G-OO/C-SO phase; t indicates the pump-probe delay (t$<$0 represents the unperturbed state). The photo-excited holons and doublons perturb locally the magnetic coupling along the FM chain $J_c^*$ ($J_c^{',*}$) (t=0, central site). The photo-excited electrons relax within few picoseconds (t$\approx$t$_1$) leaving a local perturbation on the spin system. The spin disorder diffuses along the FM chains on a longer timescale (t=t$_2$) resulting in $SW$ transfer between the HE and SP peaks (see text).}
\end{figure}

In conclusion, making use of a novel time-domain methodology, we show that the optical properties of YVO$_3$ can be rationalized introducing a bound state between a doublon and a holon, named
Hubbard exciton (HE). In this model the two oscillators at 1.8\,eV (HE) and 2.4\,eV (SP) are described as a single band originating from the same inter-Vanadium transition. Moreover, we propose that the time-domain measurements of the spectral weight exchanged between the HE and the SP band can be used to study the kinetic energy gain of the excitonic feature associated to spin order. Further studies are called to verify these new intriguing scenarios.

The authors are grateful to M. Malvestuto, J. van den Brink, L. Hozoi and K. Wohlfeld for useful discussion. We acknowledge G. R. Blake for reviewing the manuscript.

\clearpage

\begin{center}
\LARGE{\itshape{Supplementary information}}
\end{center}

\normalsize{}\section{\label{statics} Static fits}

We fitted  $\epsilon_2$ by a sum of six Gaussian peaks and a Tauc-Lorentz oscillator for the transition at lower energy (HE), $\epsilon_1$ by the Kramers-Kronig consistent functional form. With those line-shapes, that have been justified and extensively used to address amorphous and locally-disordered materials\cite{yvo:ole97,yvo:mac00}, we obtain static fits of very good quality (FIG. 7).

These are the dielectric expressions used:

$\epsilon_2(\omega) = L_a(\omega) + L_b(\omega) + \dots +L_g(\omega)$,

where $L_a(\omega) = \frac{1}{\omega}\frac{a \omega_a \gamma_a (\omega-gap)^2}{(\omega^2-\omega_a^2)^2+\omega^2\gamma_a^2}$ if $\omega > gap$ and $L_a(\omega) = 0$ elsewhere. $L_c,L_d,L_e,L_f,L_g$ are analogous of $L_b(\omega) = b(e^{-4ln2\frac{(\omega-\omega_b)^2}{\gamma_b^2}}-e^{-4ln2\frac{(\omega+\omega_b)^2}{\gamma_b^2}})$; a,b,c… are the amplitudes, $\omega_i$ and $\gamma_i$ the central frequencies and widths for $i=a,b,c$\dots

$\epsilon_1(\omega) = 1 + D_a(\omega) + D_b(\omega) + \dots + D_g(\omega)$,

where $D_a(\omega)$ has the Jellison's form\cite{yvo:jel96} and the other terms are proper sums of Dawson's functions\cite{yvo:sou06}. By simultaneous fitting of both the real and imaginary part of the dielectric constant we were able to obtain a series of continuously-varying parameters. In the c-axis fitting the Gaussian centered at 33000 cm$^{-1}$ fades below 140 K: we left this oscillator free to vary as attempts to block it produces a much worse fit. For the same reason we bind the central frequency of another oscillator at 23000 cm$^{-1}$. 
These assumptions are justified by the overall good fits of the static optical properties. 
It should be noted that, this approximations, which have been made to have a good description of the static optical properties outside of the interest range for the time domain data, do not affect the outcome of the time domains fits. This was verified by performing time domain fits with different sets of parameters.
The spectral weights have been computed as $\int_0^{23000}\omega\epsilon_2(\omega)d\omega$, with $\omega$ in $cm^{-1}$ and $\epsilon_2(\omega)$ composed only of the HE peak or the SP peak.

\begin{figure}[h!] \centering
\includegraphics[width=7.8cm]{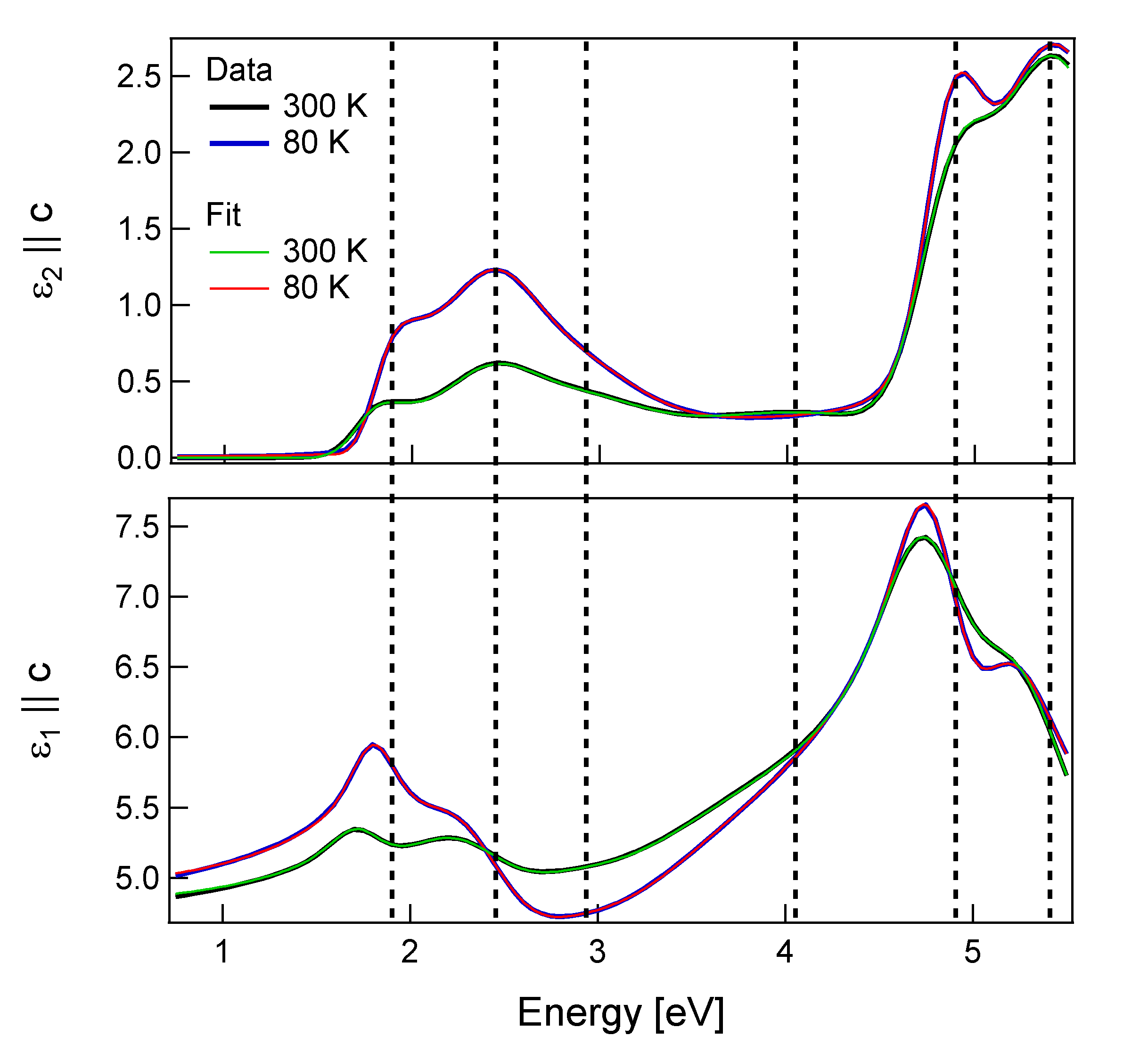}
\caption{\label{FIG1app} Imaginary (top) and real (bottom) part of the dielectric constant of YVO$_{3}$ measured by ellipsometry\cite{yvo:reu12} along the $c$ axis at 80\,K and 300\,K. The vertical dashed lines mark the positions of the central frequencies of the oscillators used to perform the static fitting procedure (see text).}
\end{figure}

\begin{figure}[t] \centering
\includegraphics[width=7.8cm]{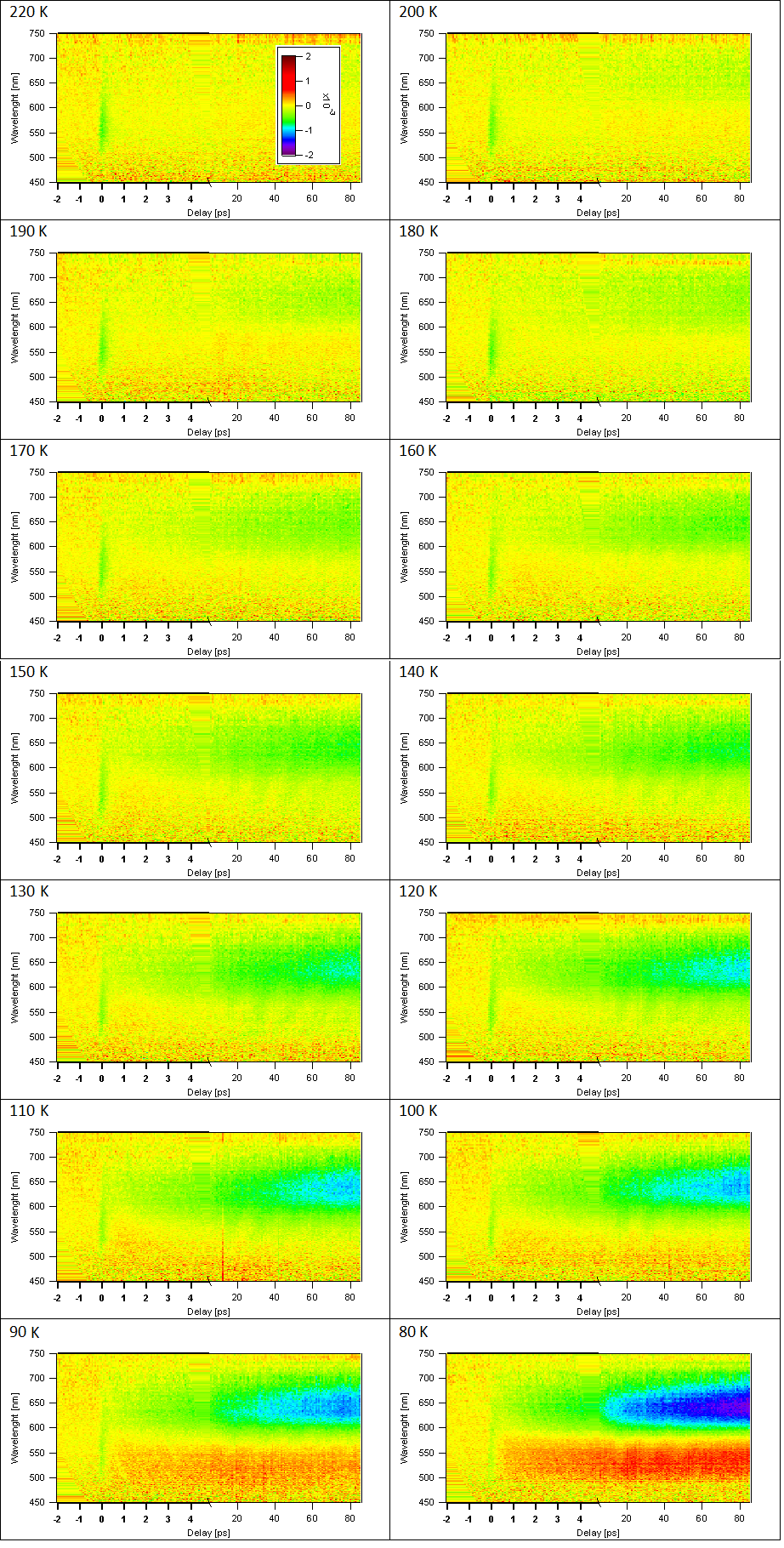}
\caption{\label{FIG2app} Relative variation of the reflectivity in the -2 to 80 ps range, with P$||$a and p$||$c.}
\end{figure}

\begin{figure}[t] \centering
\includegraphics[width=7.8cm]{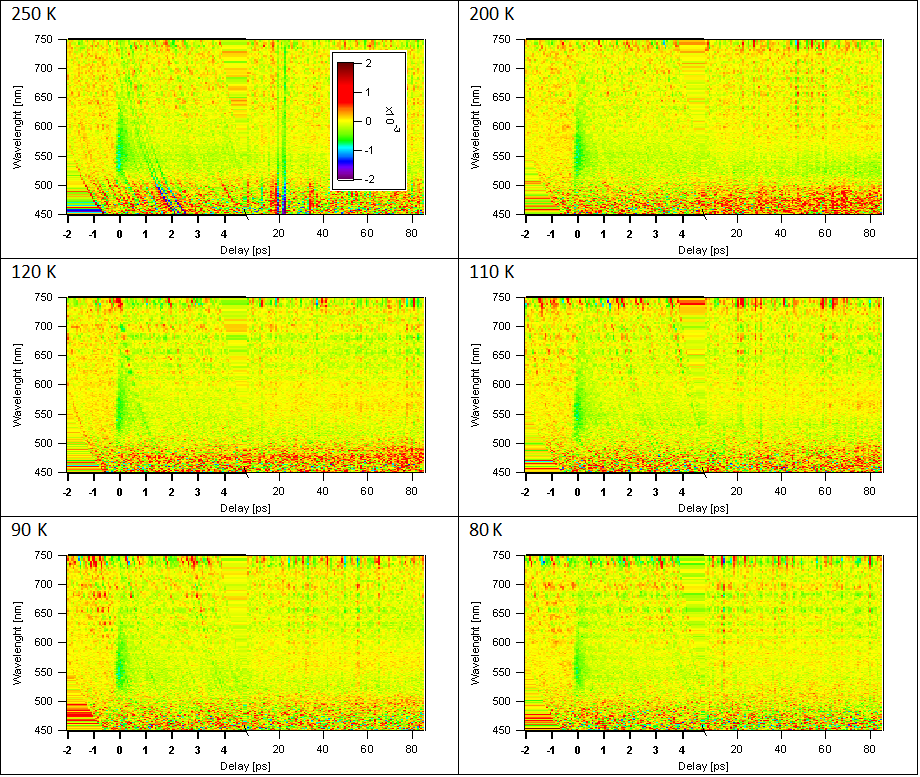}
\caption{\label{FIG3app} Measurements with P$||$c and p$||$a.}
\end{figure}

\begin{figure}[t] \centering
\includegraphics[width=7.8cm]{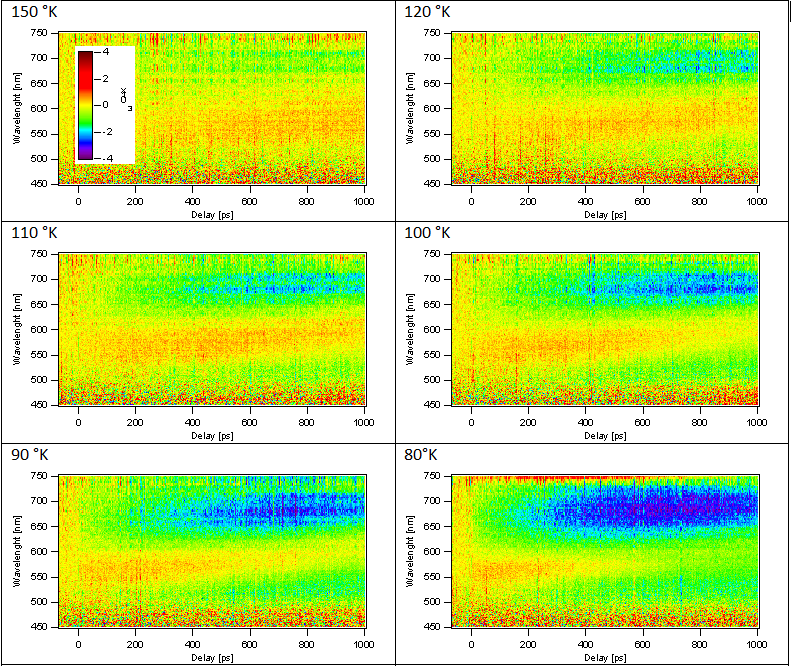}
\caption{\label{FIG4app} Relative variation of the reflectivity in the -40 to 1000 ps range, with P$||$a and p$||$c.}
\end{figure}
\section{\label{dynamics} Time-resolved measurements}

We measured the transient reflectivity $\Delta$R(t)/R as a function of temperature and pump-probe delay in the 450-750 nm wavelength-region after excitation with 4 mJ/cm$^2$ of 775 nm ultra-short Ti:Sa laser pulses at 40 KHz. The linearity of the response was checked in all phases. Time-domain measurements were performed with the pump parallel to the a-axis (P$||$a) and probe parallel to c (p$||$c) for the intervals -2$\div$4 ps and 4$\div$80 ps. A set of representative measurements is plotted on FIG. 8. Similar measurements performed with pump parallel to the c-axis (P$||$c) and probe parallel to the a-axis (p$||$a) shows only a temperature-independent fast decay time (FIG. 9). Long timescale measurements for P$||$a and p$||$c are shown in FIG. 10. The data shown are limited to temperatures as low as 80\,K, because the crystal tend to break at the low temperature phase transition often leading to a loss of thermal contact\cite{yvo:reu12} and strong average heating.

\section{\label{sw} Analysis of the spectral weight}

Starting from the fits to the static optical properties and our pump-probe data we extract the time-domain spectral weight variation $\Delta$$SW(t)$ of the HE and SP excitations as follows. 

We fit a differential model $(R(t)-R_0)/R_0$ to the measured transient reflectance $\Delta{R}_{exp}(t)/R_{exp}$, where $R_0$ and $R(t)$ are the models describing, respectively, the equilibrium reflectivity and the pump-perturbed one as a function of pump-probe delay $t$. The values of the oscillators parameters obtained by this fitting procedure at different times are used to calculate the evolution of the spectral weight:

$\Delta{SW(t)}=SW^{pumped}(t)-SW^{static}(T)$,

where $SW^{pumped}(t)$ is calculated from the fitting parameters of $R(t)$ and $SW^{static}(T)$ is obtained from the static reflectivity at temperature $T$ (both $SW$ are calculated by the numerical intergration described earlier).

\subsection{\label{thermal}Non-thermal contribution}
At any fixed temperature T, the non-thermal contribution to the variations of the SW of HE and SP can be calculated from static optical properties, the time-resolved data and the laser pump energy, as follows:

\footnotesize$\Delta{SW}^{non-thermal}(t)=SW^{pumped}(t)-SW^{static}(T+\Delta{T(t)})$\normalsize,

where $SW^{pumped}(t)$ is the photo-excited $SW$ and $SW^{static}(T+\Delta{T(t)})$ is obtained by interpolation at $T+\Delta{T(t)}$ of the static model. $\Delta{T(t)}$ is the pump-induced heating calculated from a two-temperature model (2TM)\cite{yvo:kir10,yvo:bea96} for the lattice ($L$) and spin ($S$) degrees of freedom:

$C_L\frac{dT_L}{dt}=-\gamma(T_L-T_S)+{\rho}P_{eff}(t)$

$C_S\frac{dT_S}{dt}=-\gamma(T_S-T_L)+(1-\rho)P_{eff}(t)$

where $C_L$ and $C_S$ are the heat capacities\cite{yvo:bla02} of the two subsystems, $\gamma$ is the magnetoelastic coupling and $\rho$ a phenomenological constant. In this model we assume that the pump pulse $P(t)$ photo-excites carriers from the lower Hubbard band (LHB) to the upper Hubbard band (UHB). As the quasi-particles relaxes, they act as an effective pump $P_{eff}(t)$ for the lattice and spin degrees. In this picture, $\rho$ represent the phenomelogical coupling of the electronic subsystem to the other two. The behaviour of $T_L$ and $T_S$ is reported in FIG. 11 for T=80\,K.

\begin{figure}[h!] \centering
\includegraphics[width=7.8cm]{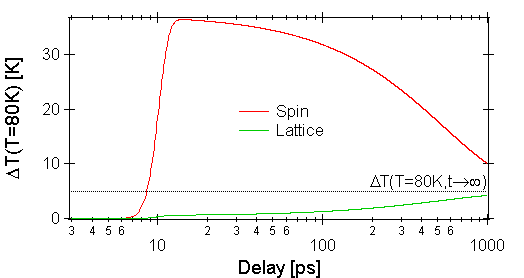}
\caption{\label{FIG5app} Two temperature model for T=80\,K. A gaussian $P_{eff}(t)$ with $FWHM=3\,ps$ is turned on at $t=10\,ps$. $T_S(t)$ is reported in red, $T_L(t)$ in green while the converging straight line $\Delta{T}(80\,K,t\rightarrow\infty)$ is dotted.}
\end{figure}

The validity of this model is confirmed by comparison, at any temperature, with the expected thermodynamic steady-state temperature increase $\widetilde{\Delta{T}(T)}$. It is straightforward to write

$\widetilde{\Delta{T}(T)}=\frac{Q_{abs}\cdot{N_A}\cdot{V}}{S\cdot{d}\cdot{u}\cdot{C_{mol}}}\approx\frac{150}{C_{mol}[J/(mol\cdot{K})]}$,

where $Q_{abs}$ is the pump energy absorbed by the sample, $N_A$ the Avogadro's number, $V$ the elementary cell volume, $S$ the sample's surface irradiated, $d$ the pump's penetration depth, $u$ the number of chemical units in a cell and $C_{mol}$ the temperature-dependent total heat capacity. 

\begin{figure}[h!] \centering
\includegraphics[width=7.8cm]{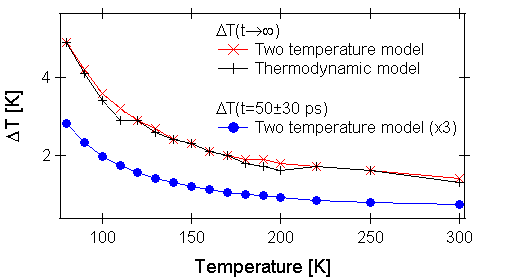}
\caption{\label{FIG6app} Comparison of the two thermodynamic models at equilibrium or long pump-probe delay. The blue dots represent the calculated lattice temperature used to calculate the non-thermal contribution to the $SW$ variations at different temperatures.}
\end{figure}

There is a good agreement between the temperature dependence of the temperature increases for the two models, as shown in FIG. 12 (red and black curves). At this point, the 2TM permits to obtain the temporal dependence of the lattice temperature and allows for the calculation of the non-thermal component. The blue dots in FIG. 12 represent the temperature variations at pump-probe delay $t=50\,ps$ used to obtain the non-thermal contribution to the variation of the spectral weight (FIG. 5).

\end{document}